\documentclass[11pt]{article}

\usepackage{moriond,epsfig}

\newcommand{\Dz}{\ensuremath{D^0}}

\newcommand{\Bzbar}{\ensuremath{\overline{B}{}^0}}

\newcommand{\Xa}{\ensuremath{X(3872)}}
\newcommand{\Zchargeda}{\ensuremath{Z^+(4430)}}
\newcommand{\Zchargedb}{\ensuremath{Z^+(4050)}}
\newcommand{\Zchargedc}{\ensuremath{Z^+(4250)}}

\newcommand{\Ya}{\ensuremath{Y(4008)}}
\newcommand{\Yb}{\ensuremath{Y(4260)}}
\newcommand{\Yc}{\ensuremath{Y(4360)}}
\newcommand{\Ycx}{\ensuremath{Y(4325)}}
\newcommand{\Yd}{\ensuremath{Y(4660)}}

\newcommand{\BB}{\ensuremath{B\overline{B}}}

\newcommand{\epem}{\ensuremath{e^+ e^-}}
\newcommand{\cc}{\ensuremath{c\overline{c}}}

\newcommand{\gisr}{\ensuremath{\gamma_\mathrm{ISR}}}

\newcommand{\BR}{\ensuremath{\mathcal B}}

\def\Journal#1#2#3#4{{#1} {\bf #2}, #3 (#4)}
\def\ARXIV#1#2{arXiv:{#2} [#1]}
\def\ARXIVOLD#1#2{arXiv:#1/{#2}}

\def\NIMA{{\em Nucl. Instrum. Methods} A}

\def\PLB{{\em Phys. Lett.} B}
\def\PRL{\em Phys. Rev. Lett.}
\def\PRD{{\em Phys. Rev.} D}

\def\be{\begin{equation}}
\def\ee{\end{equation}}
\def\bea{\begin{eqnarray}}
\def\eea{\end{eqnarray}}

\usepackage{relsize} 
\def\babarsym{\mbox{\slshape B\kern-0.1em{\smaller A}\kern-0.1em
    B\kern-0.1em{\smaller A\kern-0.2em R}}}
\graphicspath{{figs/}}

\title{%
  \uppercase{Charmonium[-like] states at Belle and \babarsym }
}

\author{%
  \uppercase{M. Bra\v cko}
}

\address{%
  University of Maribor, Smetanova ulica 17, SI-2000 Maribor, Slovenia\\
  and\\
  Jo\v zef Stefan Institute, Jamova cesta 39, SI-1000 Ljubljana, Slovenia
}

\begin{document}
\vspace*{4cm}

\maketitle

\abstracts{%
  Belle and \babarsym\ experiments at the KEKB and PEP-II
  $B$-factories provide also an excellent environment for spectroscopy
  studies. In this report we present recent results in the field of
  charmonium spectroscopy, focusing on new charmonium-like states
  observed in $B$ decays, and on $J^{PC}=1^{--}$ resonances created in
  $\epem$ annihilation through the photon radiative return.
}

\section{Introduction}

There has been a renewed interest in charmonium spectroscopy since
2002. The attention to this field was directed by the discovery of the
two missing $\cc$ states below the open-charm threshold, $\eta_c (2S)$
and $h_c (1P)$,\cite{ref:etac_2S_discovery,ref:hc_1P_discovery} but
even more by observations of a number of new
particles~\cite{ref:swanson} above the threshold for the open-charm
production. Many of these exciting new states -- although resembling
charmonia -- differ from regular $\cc$ states by some of their
properties, or can simply not be identified as charmonia due to lack
of available $\cc$ assignments. The naming convention, $X$, $Y$, $Z$, 
already indicates lack of knowledge about the structure and properties
of these new states at the time of their discovery.

The majority of these new states were observed by the
Belle~\cite{ref:Belle} and \babarsym~\cite{ref:Babar} experiments,
operating their detectors at the two respective asymmetric-energy
$\epem$ colliders (so-called \emph{B-factories}): KEKB in Japan and
PEP-II in the USA. Both experiments together have by now accumulated
huge data samples that in total correspond to nearly
$1.5~\mathrm{ab^{-1}}$ and contain $1.3 \times 10^{9}$ \BB\
pairs.\footnote{PEP-II was turned off in April 2008 and the final
\babarsym\ data sample corresponds to $531~\mathrm{fb^{-1}}$.}
Although initially designed for measurements of $CP$ violation in the
$B$-meson system, experiments at $B$-factories can also use large
samples of experimental data to perform searches for new states and to
study their properties. The charmonium(-like) particles are at
$B$-factories produced by several mechanisms: via $B$ decays; in
$\epem$ annihilation into double $\cc$; $C$-even states can be formed
in $\gamma\gamma$ processes; and $J^{PC}=1^{--}$ resonances can be
created in $\epem$ annihilation after the photon radiative return. In
this review we will only present results from recent analyses of new
states produced by the first and the last of the four mentioned
mechanisms.

\section{Charmonium-like states observed in $B$ decays}

The story about new charmonium-like states begins in 2003, when Belle
reported~\cite{ref:X3872_belle_discovery} on the $B^+ \to K^+ J/\psi
\pi^+\pi^-$ analysis,\footnote{In this review, the inclusion of
charge-conjugated states is always implied.} where a new state
decaying to $J/\psi \pi^+\pi^-$ was discovered, and soon
confirmed~\cite{ref:X3872_confirm} by the CDF, D\O\ and \babarsym\
collaborations. The narrow \Xa\ state~\cite{ref:pdg2008} ($\Gamma =
(3.0^{+1.9}_{-1.4}\pm 0.9)~\mathrm{MeV}$), with a mass of
$(3872.3\pm0.8)~\mathrm{MeV}/c^2$ very close to the $\Dz
\overline{D}{}^{\ast 0}$ threshold, was studied intensively in several
production and decay modes by Belle, \babarsym\ and other
experiments.\cite{ref:X3872_ref1,ref:X3872_ref2,ref:X3872_ref3,ref:X3872_ref4,ref:X3872_ref5,ref:X3872_ref6,ref:X3872_ref7,ref:X3872_ref8}
These results suggest two possible $J^{PC}$ assignments, $1^{++}$ and
$2^{-+}$, and establish the \Xa\ as a candidate for a loosely bound $\Dz
\overline{D}{}^{\ast 0}$ molecular state. However, results
provide substantial evidence that the \Xa\ state must contain a
significant \cc\ component as well. A much larger data sample is
probably required to resolve this issue.

\subsection{Charged charmonium-like states: \Zchargeda; 
\Zchargedb \& \Zchargedc}

\begin{figure}[t]
  \begin{center} 
    \begin{minipage}[b]{0.48\textwidth}%
      \centerline{
        \begin{tabular}{c}
          \includegraphics[width=0.5\textwidth]
          {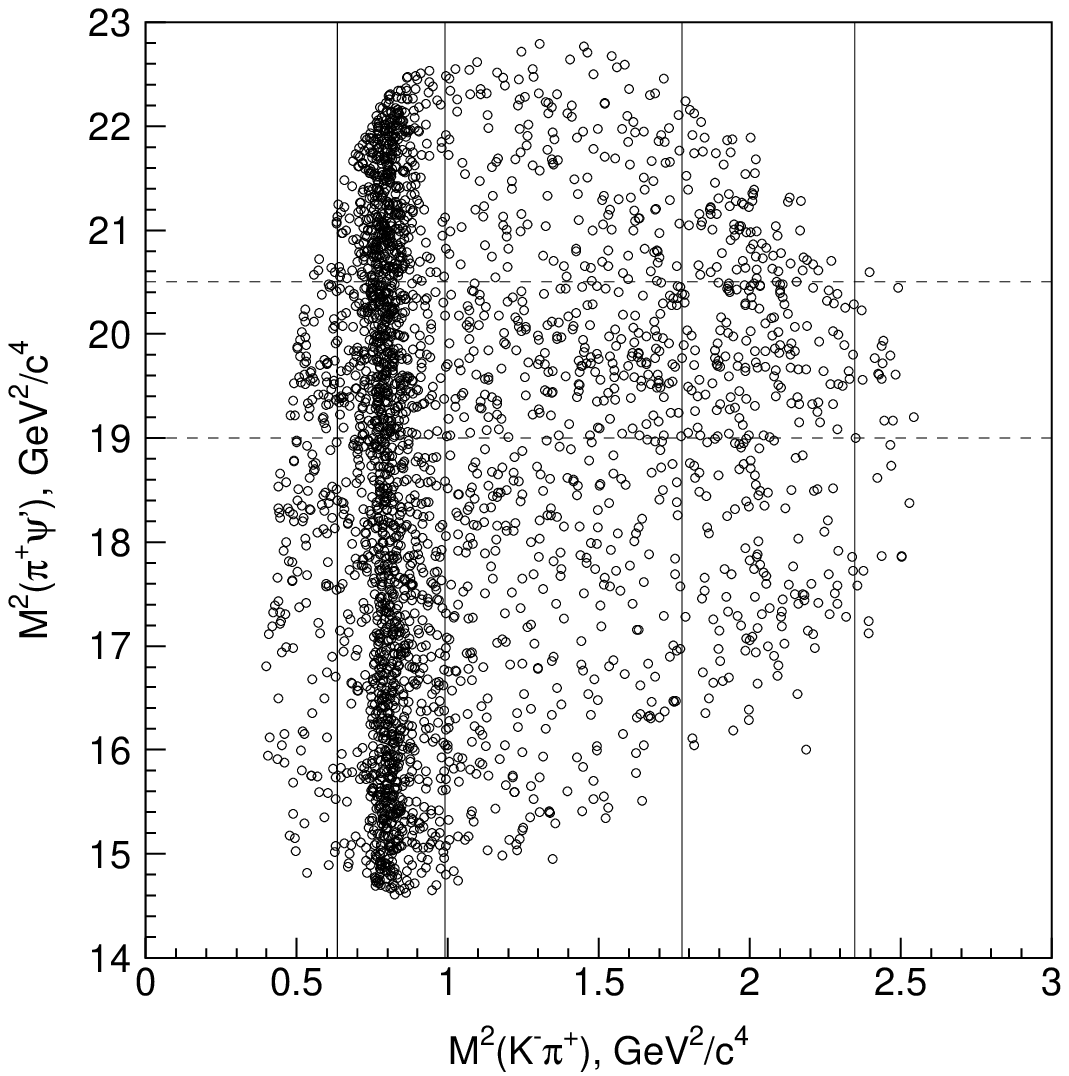}        \\
          \includegraphics[width=0.5\textwidth]
          {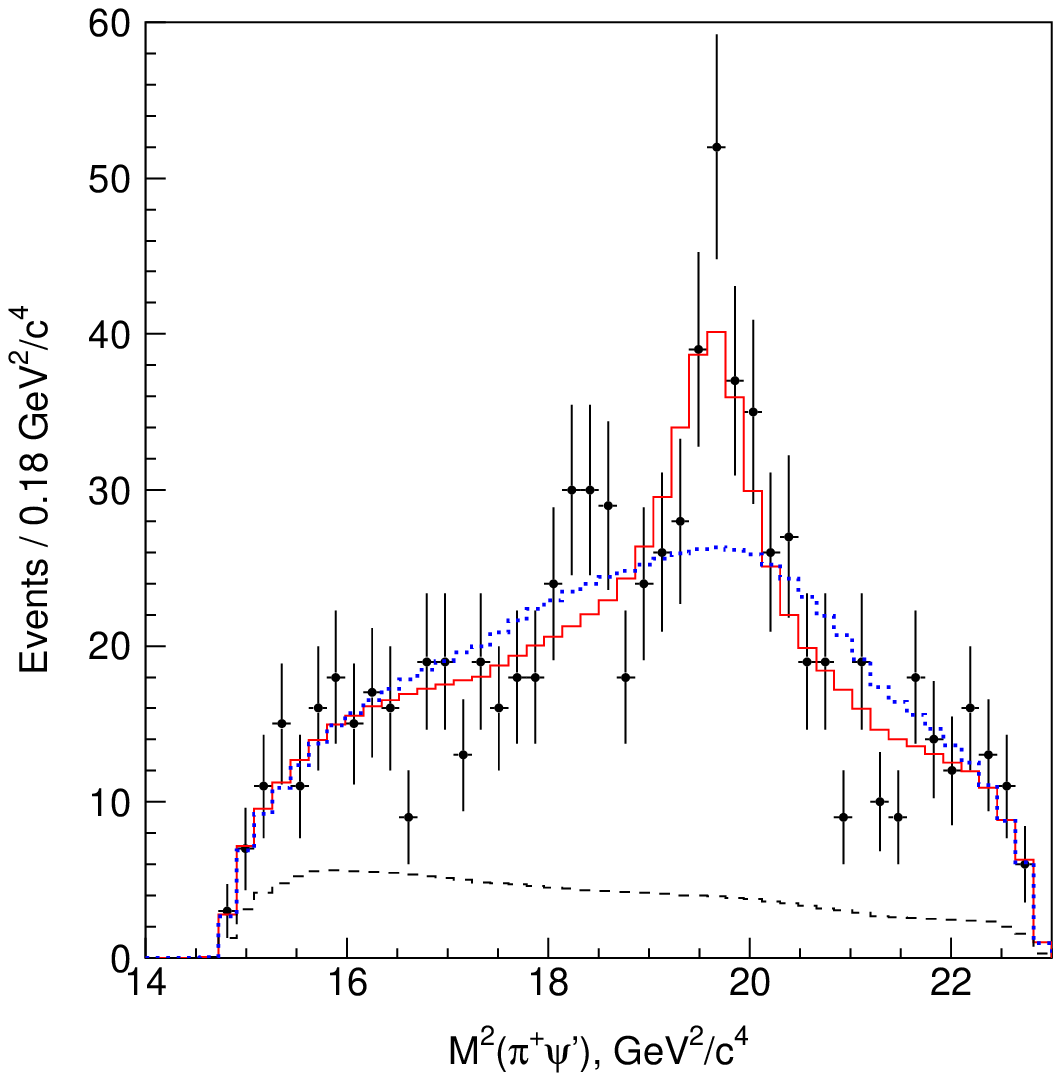}  
        \end{tabular}
        \begin{tabular}{c}
          \includegraphics[width=0.55\textwidth]
          {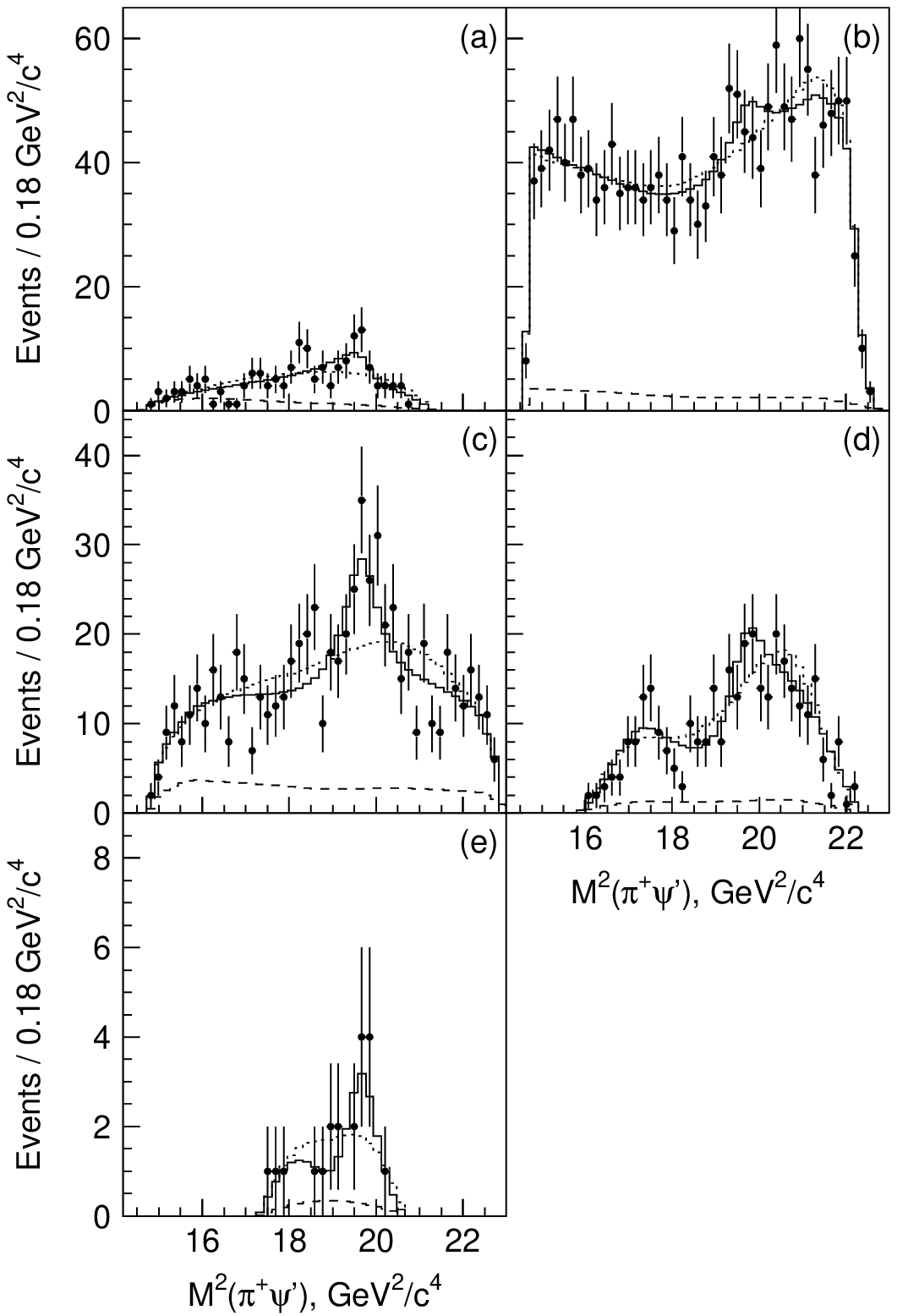}
        \end{tabular}
      }
      \caption{%
        [Belle results]
        \textbf{Top left}:The $B \to K \pi^+ \psi(2S)$ Dalitz
        plot. The second and the fourth of five vertical slices
        correspond to the $K^\ast(890)$ and $K^\ast(1430)$ regions, 
        respectively. \textbf{Bottom left}: The Dalitz plot projection
        for the $\pi^+ \psi (2S)$ invariant mass with the $K^\ast$ 
        veto applied. The solid (dotted) histogram shows the fit result with a
        single (without any) $\pi^+ \psi (2S)$ state. \textbf{Right
          panel}: (a)-(e) plots show the fit results for five
        vertical slices of the Dalitz plot from left to right.
        \label{fig:z4430_plots_belle}
      }
    \end{minipage} 
    \hfill
    \begin{minipage}[b]{0.5\textwidth}%
      \centerline{
        \begin{tabular}{c}
          \includegraphics[width=1.0\textwidth]
          {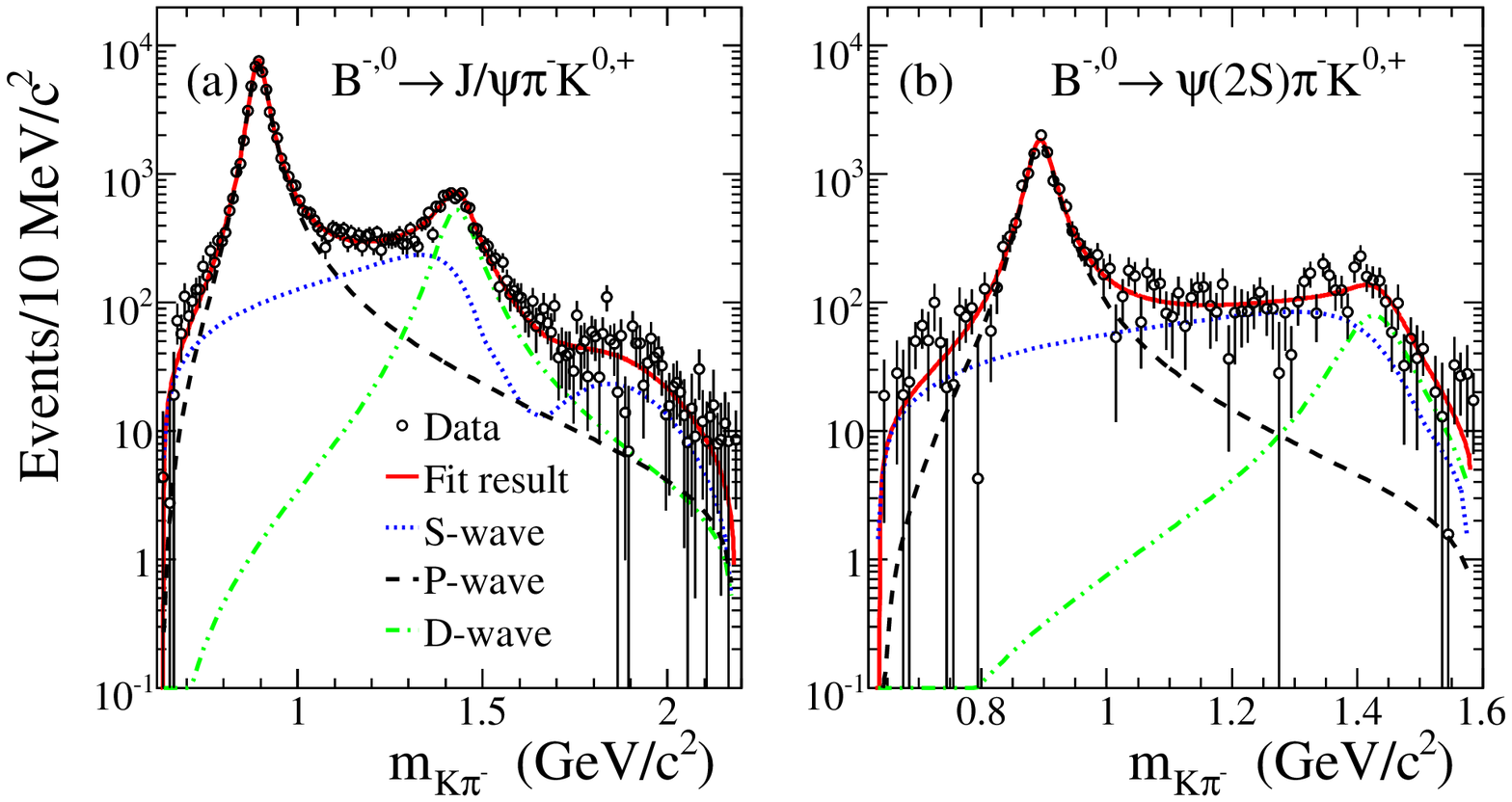}\\
          \includegraphics[width=1.0\textwidth]
          {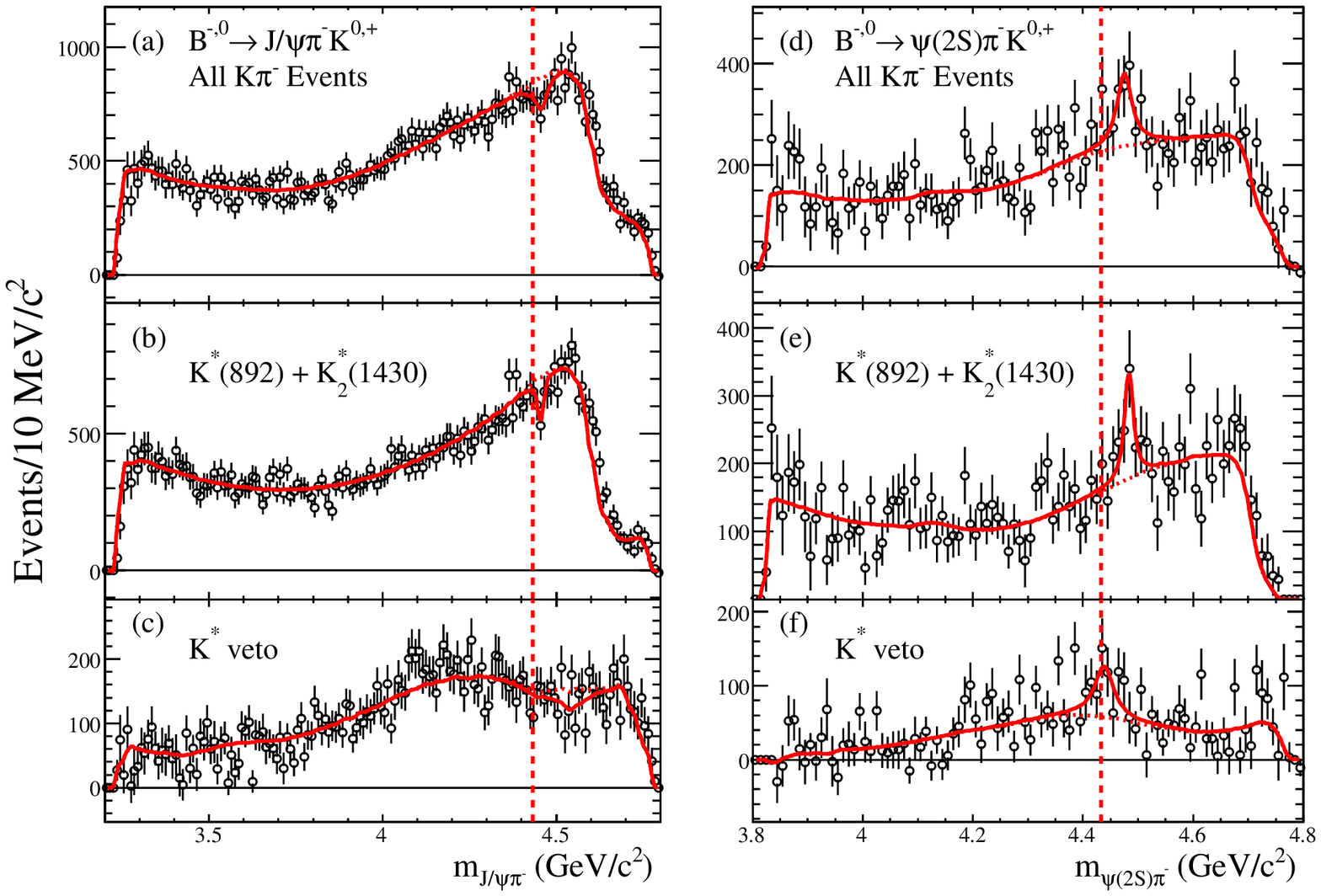}
        \end{tabular}
      }
      \caption{%
        [\babarsym\ results] 
        \textbf{Top}: The fit results to the $K \pi^-$ mass
        distributions for the ((a),(b)) $B^{-,0} \to \pi^-
        K^{0,+} (J/\psi , \psi (2S)) $ decays. 
        \textbf{Bottom panel}: The $J/\psi \pi^-$ 
        (left) and $\psi (2S) \pi^-$ (right) mass distributions for
        all events, and events inside or outside the $K^\ast$ regions.
        \label{fig:z4430_plots_babar}
      }
      \end{minipage}
    \end{center}
\end{figure}
Last year a surprising discovery of a new charmonium-like state was
reported~\cite{ref:Z4430_belle_discovery} by Belle in the $B^{+,0} \to
K^{0,-} \pi^+ \psi(2S)$ analysis, performed on a data sample with $657
\cdot 10^6$ \BB\ pairs. After excluding the $K \pi$ Dalitz regions
that correspond to $K^\ast (890)$ and $K_2^\ast (1430)$ mesons
(\emph{i.e.} $K^\ast$ veto), a strong enhancement is seen in the
$\pi^+ \psi (2S)$ invariant mass distribution. A fit with a
Breit-Wigner shape yields a peak mass and width of $M = (4433 \pm 4
\pm 2)~\mathrm{MeV}/c^2$ and $\Gamma =
(45^{+18}_{-13}{}^{+30}_{-13})~\mathrm{MeV}$, with a $6.5\sigma$
statistical significance. The observed resonance, called \Zchargeda,
is the first charged charmonium-like meson state -- an obvious
tetraquark candidate.

Using the same data sample as above, Belle also performed a full
Dalitz plot analysis~\cite{ref:Z4430_belle_dalitz} with a fit model
that takes into account all known $K \pi$ resonances below
$1780~\mathrm{MeV}/c^2$. Dalitz plot is divided in five $M^2(K \pi)$
regions and the $\Zchargeda$ signal is clearly seen for the
$K^\ast$-veto-equivalent $M^2(\pi^+ \psi (2S))$ distribution,
\emph{i.e.} for the sum of the $1^\mathrm{st}$, $3^\mathrm{rd}$ and
$5^\mathrm{th}$ regions (see Fig.~\ref{fig:z4430_plots_belle}). The
fit results with $6.4\sigma$ peak significance agree with previous Belle 
measurement, and provide the updated $\Zchargeda$ parameters: $M =
(4443^{+15}_{-12}{}^{+19}_{-13})~\mathrm{MeV}/c^2$, $\Gamma =
(109^{+86}_{-43}{}^{+74}_{-56})~\mathrm{MeV}$ and $\BR (\Bzbar \to K^-
\Zchargeda ) \times \BR (\Zchargeda \to \pi^+ \psi (2S) ) =
(3.2^{+1.8}_{-0.9}{}^{+5.3}_{-1.6}) \cdot 10^{-5}$.

\babarsym\ also searched for the $\Zchargeda$ signature in their data
sample, analysing the $B^{-,0} \to \psi \pi^- K^{0,+}$ ($\psi$ =
$J/\psi$ or $\psi (2S)$) decays.\cite{ref:Z4430_babar_search} A
substantial amount of work in this analysis is invested into a
detailed study of the $K \pi^-$ system, since its mass and
angular-distribution structures strongly influence the Dalitz
plots. As shown in Fig.~\ref{fig:z4430_plots_babar} the $K \pi^-$
invariant mass distributions are well described in terms of a
superposition of $S-$, $P-$ and $D-$wave amplitudes. The shapes and
the composition of these components strongly affect the $\psi \pi^-$
mass spectrum through the $K \pi^-$ reflection background. However, it
is found that these reflections alone can not explain a narrow peak in
the $J/\psi \pi^-$ or $\psi (2S) \pi^-$ mass distributions. These
distributions for all events, and separately for events inside and
outside the $K^\ast(890)$ and $K^\ast(1430)$ regions, are then fitted
with the sum of the $K \pi^-$ background function and a relativistic 
Breit-Wigner shape (see Fig.~\ref{fig:z4430_plots_babar}). No
significant evidence for a signal peak is seen for any of the
processes investigated, not even in the $K^\ast$ veto region for the 
$\psi (2S) \pi^+$ distribution, where the $\Zchargeda$ was observed 
by Belle. The most prominent structure in the $\psi (2S) \pi^-$ mass 
distribution for all events is an excess of $2.7\sigma$ with a mass
and width of $M = (4476 \pm 8(\mathrm(stat.))~\mathrm{MeV}/c^2$ 
and $\Gamma = 32\pm 16(\mathrm(stat.))~\mathrm{MeV}$. Using the 
Belle values~\cite{ref:Z4430_belle_discovery} for the $\Zchargeda$, the
upper limit for the product of branching fractions is calculated as
$\BR (\Bzbar \to K^-\Zchargeda ) \times \BR (\Zchargeda \to \pi^+ \psi
(2S) ) < 3.1 \cdot 10^{-5}$ at a $95\%$ confidence level. This 
result gives no conclusive evidence for the existence of the $\Zchargeda$,
seen by Belle.

The observation of the $\Zchargeda$ state suggests that studies of $B
\to K \pi (\cc)$ decays could reveal other similar neutral and charged
partners. Belle thus reports also on a Dalitz plot analysis of $\Bzbar
\to K^- \pi^+ \chi_{c1}$ decays with $657 \cdot 10^6$ \BB\
pairs.\cite{ref:Z1Z2_belle_discovery} The fit model for $K \pi$
resonances is the same as in the $\Zchargeda$ Dalitz analysis, but
here it includes also the $K_3^\ast(1780)$ meson. The fit results
suggest that a broad doubly peaked structure in the $\pi^+ \chi_{c1}$
invariant mass distribution should be interpreted by two new states,
called $\Zchargedb$ and $\Zchargedc$. The double-$Z^+$ hypothesis is
favoured when compared to the single-$Z^+$ (no-$Z^+$) hypothesis by
the statistical significance of $5.7\sigma$ ($13.2\sigma$), and even
with various systematic variations of the fit model, the significance
is still at least $5.0\sigma$ ($8.1\sigma$). The masses, widths and
product branching fractions for the two states are: $M(\Zchargedb) =
(4051 \pm {14} {}^{+20}_{-41})~\mathrm{MeV}/c^2$, $\Gamma(\Zchargedb)
= (82^{+21}_{-17}{}^{+47}_{-22})~\mathrm{MeV}$, $M(\Zchargedc) =
(4248^{+44}_{-29} {}^{+180}_{-35})~\mathrm{MeV}/c^2$,
$\Gamma(\Zchargedc) = (177^{+54}_{-39}{}^{+316}_{-61})~\mathrm{MeV}$;
and $\BR(\Bzbar \to K^- \Zchargedb ) \times \BR (\Zchargedb \to \pi^+
\chi_{c1} ) = (3.0^{+1.5}_{-0.8}{}^{+3.7}_{-1.6}) \cdot 10^{-5}$,
$\BR(\Bzbar \to K^- \Zchargedc ) \times \BR (\Zchargedc \to \pi^+
\chi_{c1} ) = (4.0^{+2.3}_{-0.9}{}^{+19.7}_{-0.5}) \cdot 10^{-5}$.

\subsection{Studies of $J^{PC}=1^{- -}$ states using ISR}

The annihilation through initial-state radiation (ISR), $\epem \to
\gisr X_\mathrm{final}$, has proven to be a powerful tool to search
for $1^{- -}$ states at $B$-factories: it enables a scan across a
broad $\sqrt{s}$ energy range below the initial $\epem$ centre-of-mass
(CM) energy, while the high luminosity compensates for the suppression
due to the hard-photon emission. The ISR processes are effectively 
identified by a small mass recoiling against the studied
system $X_\mathrm{final}$. \babarsym\ used this technique for a
discovery of the $\Yb$ state above $D^{(\ast)}\overline{D}{}^{(\ast)}$
threshold in the $\epem \to \gisr \Yb \to \gisr J/\psi \pi^+ \pi^-$
process.\cite{ref:Y4260_babar_discovery} Using the same method Belle
recently confirmed~\cite{ref:Y4260_belle_confirm} the $\Yb$ state, but
also found another resonant structure, called $\Ya$. A similar
analysis was performed by Belle to study the ISR $\epem $ annihilation
process resulting in the $\psi(2S) \pi^+ \pi^-$ final
state.\cite{ref:Y4360_Y4660_belle_discovery} The obtained $\psi(2S)
\pi^+ \pi^-$ mass distribution reveals two resonant structures, called
$\Yc$ and $\Yd$. While $\Yd$ still needs a confirmation, the former
resonance, $\Yc$, has a mass similar to the wide structure $\Ycx$,
discovered previously by \babarsym.\cite{ref:Y4360_babar_discovery} 

Results are summarised in Table~\ref{tab:Y_properties}. The $Y$ states
observed in $J/\psi \pi^+\pi^-$ and $\psi(2S) \pi^+ \pi^-$ decay modes
are distinctive, although a hint exists that the $\Yb$ could also be
seen in the $\psi(2S) \pi^+\pi^-$ decay
mode.\cite{ref:YISR_pipipsi2s_combined_fit} The nature of $Y$ states
and their strong couplings to $\psi \pi^+ \pi^-$ are somewhat
puzzling: such heavy charmonium(-like) states should decay mainly to
$D^{(\ast)}\overline{D}{}^{(\ast)}$, but it seems that observed $Y$
states do not match the peaks in $\epem \to D^{(\ast)\pm}
D^{(\ast)\mp}$ cross sections, measured by
Belle.\cite{ref:dstdstbar_csx_isr_belle} This conclusion is supported also by \babarsym\
cross-section measurements.\cite{ref:ddbar_csx_isr_babar,ref:dstdstbar_csx_isr_babar}

\begin{table}[t]

  \begin{center}
    \caption{Properties of $Y( 1^{--})$ states, measured by Belle 
      and \babarsym . States marked with (?) might be the same.}
  \label{tab:Y_properties}
    \begin{tabular}{|cccccc|} \hline
                                  &             
      & \multicolumn{2}{c}{Belle} 
      & \multicolumn{2}{c|}{\babarsym}                         \\ \hline 
      Y state                     & Decay mode 
      & $M$ ($\mathrm{MeV}/c^2$)& $\Gamma$ ($\mathrm{MeV}$) 
      & $M$ ($\mathrm{MeV}/c^2$)& $\Gamma$($\mathrm{MeV}$) \\ \hline
      \Ya                         & $J/\psi \pi^+ \pi^-$ 
      & $4008 \pm 40 {}^{+114}_{-28}$ & $226 \pm 44 \pm 87 $ 
      &                                &                        \\
      \Yb                              & $J/\psi \pi^+ \pi^-$ 
      & $4247 \pm 12{}^{+17}_{-32}$ & $108 \pm 19 \pm 10$ 
      & $4259 \pm 6 {}^{+2}_{-3}$   & $105 \pm 18 {}^{+4}_{-6}$ \\ 
      \Ycx (?)                          & $\psi(2S) \pi^+ \pi^-$ 
      &                                & 
      &  $4324 \pm 24$                & $172 \pm 33 $         \\
      \Yc (?)                          & $\psi(2S) \pi^+ \pi^-$ 
      & $4361 \pm 9 \pm 9$           & $74 \pm 15 \pm 10 $ 
      &                                &                        \\
      \Yd                              & $\psi(2S) \pi^+ \pi^-$ 
      & $4664 \pm 11 \pm 5$          & $48 \pm 15 \pm 3$
      &                                &                         \\ \hline
    \end{tabular} 
  \end{center}
\end{table}

\section{Summary and Conclusions}

The $B$-factory experiments, Belle and \babarsym , provide an excellent
environment for charmonium spectroscopy. As a result, many new
charmonium-like particles have been discovered during their
operation, and some of them -- like \Xa , \Zchargeda, \Zchargedb\ \&
\Zchargedc and several $Y(1^{--})$ states -- are mentioned in this report.

\section*{References}


\end{document}